\pdfoutput=1
\documentclass[prd,aps,nofootinbib,twocolumn,superscriptaddress,
preprintnumbers,balancelastpage,longbibliography]{revtex4-1}
\usepackage{ae,aecompl}
\usepackage{graphicx}	
\usepackage{amsmath}	
\usepackage{graphicx}
\usepackage{dcolumn}
\usepackage{bm}
\usepackage{graphics}
\usepackage{afterpage}
\usepackage{float}
\usepackage{rotating}
\usepackage{multirow}
\usepackage{tabularx}
\usepackage{booktabs}
\usepackage{multirow}
\usepackage{fancyhdr}
\usepackage{hyperref}
\usepackage[utf8]{inputenc}
\usepackage{theorem}
\usepackage{moreverb}
\usepackage{euscript}
\usepackage{psfrag}
\usepackage{slashed}
\usepackage{mathtools}
\usepackage{makecell}
\usepackage[flushleft]{threeparttable}
\usepackage[english]{babel}
\usepackage{bm}
\usepackage[mathlines]{lineno}
\usepackage{color}
\usepackage[normalem]{ulem}
\definecolor{rossoCP3}{cmyk}{0,.88,.77,.40}
\definecolor{darkred}{rgb}{0.6,0,0}
\definecolor{drkgrn}{RGB}{0, 51, 0}
\hypersetup{
     colorlinks   = true,
     citecolor    = blue,
     urlcolor     = blue,
      linkcolor    = darkred
}

\usepackage{listings}
\usepackage{color,xcolor}

\newcommand{\be}{\begin{equation}}

\newcommand{\ee}{\end{equation}}
\newcommand{\bea}{\begin{eqnarray}}
\newcommand{\eea}{\end{eqnarray}}
\newcommand{\nuclide}[2]{{}^{#1}\mathrm{#2}}

\renewcommand{\arraystretch}{1.8}
\newcolumntype{C}[1]{>{\centering\let\newline\\\arraybackslash\hspace{0pt}}m{#1}}

\begin{document}
\title{Resonant Induced Orbital Electron Capture: Novel method for 
low-energy $\bar{\nu}_e$ detection}

\author{Evgeny Akhmedov}
\email{akhmedov@mpi-hd.mpg.de}
\author{Thierry Lasserre}
\email{thierry.lasserre@mpi-hd.mpg.de}
\author{Leonardo Maturi}
\email{leonardo.maturi@mpi-hd.mpg.de}

\affiliation{Max-Planck-Institut f\"{u}r Kernphysik, Saupfercheckweg 1, 69117 
Heidelberg, Germany}

\date{\today}

\newcommand{\mk}[1]{{\bf #1}}
\newcommand{\om}[1]{\textcolor{red}{#1}}
\newcommand{\sh}[1]{\textcolor{blue}{#1}}

\begin{abstract}
We propose a novel approach to detecting low-energy electron 
antineutrinos based on the induced capture of orbital electrons by 
nuclei. This process is resonant, requiring the antineutrino energy to 
precisely match the energy difference between the final and initial 
atomic systems. For continuous-spectrum sources, the resonance conditions 
can be satisfied without fine-tuning, and the effective cross sections 
depend on the spectral intensity of the $\bar{\nu}_e$ flux at the 
resonance rather than on the neutrino energy itself. This opens the 
possibility of detecting neutrinos of never previously probed low energies. 
We identify a number of candidate nuclides that allow transitions to excited 
states of the daughter nuclei corresponding to resonant 
$\bar{\nu}_e$ energies below the inverse $\beta$ decay threshold of 1.8\,MeV,  
and we consider a distinctive atomic–nuclear coincidence signature for 
background rejection. We discuss implications of the proposed method 
for detecting low-energy reactor neutrinos, geoneutrinos, and  
keV-scale thermal solar neutrinos. Applications to neutrino oscillation 
experiments and to reactor monitoring are also briefly discussed. 
\end{abstract}

\maketitle

{\textit{Introduction---}}To date, all experiments on neutrino detection  
have been based on just two types of underlying processes:  
(i) charged-current-mediated capture on free protons 
or nuclei with production of the corresponding charged leptons, and (ii) 
scattering off electrons, nucleons, or nuclei. 
The scattering processes are 
mediated by weak neutral currents or, in the case of $\nu_e$ or 
$\bar{\nu}_e$ scattering off electrons, by both charged and neutral currents 
(see, e.g., \cite{Formaggio:2012cpf}). 
In the present Letter, we propose yet another approach to the detection of 
electron antineutrinos -- resonant capture. The underlying reaction is  
\be
\bar{\nu}_e+[e^-+(Z,A)] \to (Z-1,A)\,,
\label{eq:rioec1}
\ee
where the square brackets indicate that the $e^-$ is a bound atomic electron. 
In this process, the incident $\bar{\nu}_e$ induces the capture of  
an electron from one of the low-lying atomic orbitals of the parent atom by 
the nucleus it hosts.  Since the final state is a single body, the process is 
resonant: it can only occur if the energy of the incoming neutrino%
\footnote{Throughout this Letter we discuss electron antineutrinos; for 
brevity we will often refer to them simply as antineutrinos or neutrinos.} 
precisely matches the energy difference between the daughter and parent 
atoms, including the excitation energy of the daughter system 
[see Eq.~(\ref{eq:Eres1})]. The process can 
thus be called resonant induced orbital electron capture (RIOEC). 

This reaction is a cross-channel of the well-known 
spontaneous electron capture (EC) process, 
$[e^-+(Z,A)] \to (Z-1,A)+\nu_e$ \cite{Bambynek:1977zz}. In both 
EC and RIOEC, the final-state atom is always produced  
in an excited state because it has a vacancy in one of the low-lying electron 
orbitals. The atom then promptly de-excites by emitting $X$-rays and/or Auger 
electrons \cite{Bambynek:1977zz,Bambynek:1972eai}.   

The possibility of RIOEC was first mentioned by Fermi    
\cite{Fermi1950nuclear}. It has subsequently been discussed by 
a number of authors, mainly in connection with 
the idea of resonance-fluorescence-type experiments with neutrinos. In such 
experiments, monoenergetic neutrinos would first be produced in nuclear 
bound-state $\beta$ decay and then detected in the inverse process, 
which is RIOEC. The recoil energies of the emitter and absorber atoms, which 
destroy the resonance, would then be either suppressed through a  
M\"{o}ssbauer-type effect \cite{Visscher:1959kka,Kells:1982rm,Kells:1983iac,
Raghavan:2005gn,Potzel:2006ad,Akhmedov:2008jn} or compensated by boosting the 
neutrino source \cite{Oldeman:2009wa}. RIOEC has also been discussed in 
connection with reactor neutrino experiments \cite{Mikaelyan:1967a}, the 
detection of geoneutrinos \cite{Marx:1969be,Krauss:1983zn} and of neutrinos 
from the cosmic neutrino background (C$\nu$B) 
\cite{Lazauskas:2007da,Cocco:2009rh,Lusignoli:2010eq,Vergados:2013qpa,Lee:2018boo}. 
The suggested RIOEC observation methods were the detection of 
$X$-rays and Auger electrons from the de-excitation of the final-state atoms 
and, when the daughter nuclei are unstable, the detection of their decay 
by radiochemical methods. 
It has generally been considered that such experiments would be 
extremely difficult, and so far RIOEC has not been observed. 

In the present Letter we propose a realistic and practical method for 
detecting RIOEC: searching for transitions to excited states of the daughter 
nuclei, in which either the excited nuclear state directly populated by 
RIOEC is  
relatively long-lived ($\tau\gtrsim 10$\,ns), or the daughter nucleus  
passes through one or more such long-lived excited states in the course 
of its de-excitation. The RIOEC process can then be detected by observing 
delayed coincidences between $X$-rays/Auger electrons from atomic 
de-excitation and $\gamma$-rays/internal conversion (IC) electrons produced 
in nuclear de-excitation. 
\begin{figure*}[t]
\centering
\includegraphics[width=8.4cm]{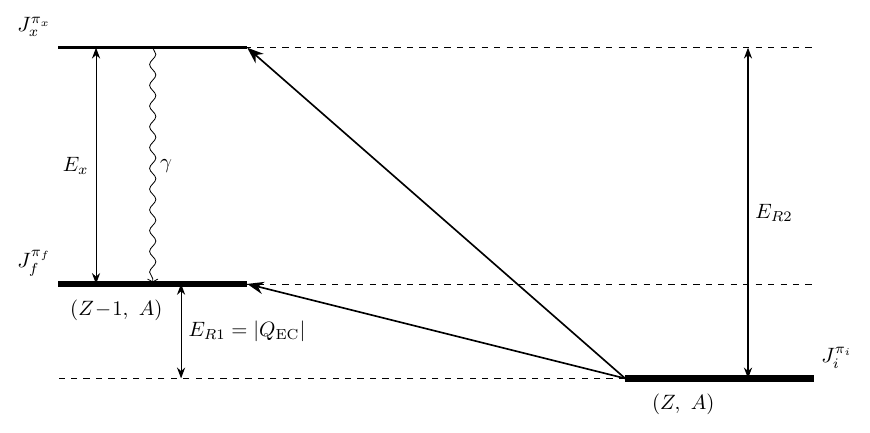}
\hspace*{4.0mm}
\includegraphics[width=8.8cm]{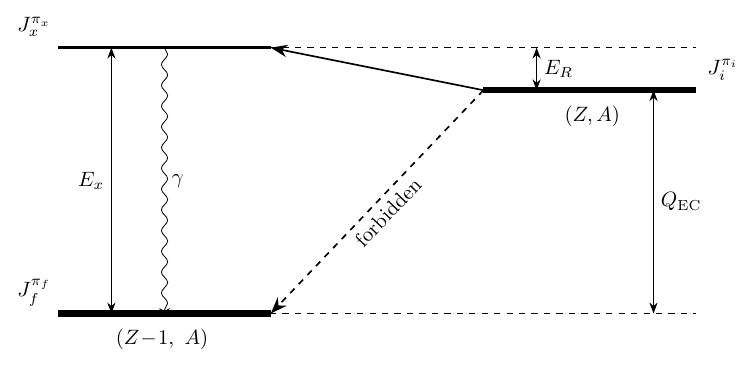}
\caption{Simplified transition schemes for RIOEC. Left panel: stable 
parent nucleus. Possible induced transitions to the ground state 
and to an excited state of the daughter nucleus are shown. Right panel: 
RIOEC transition 
in the case of an unstable target nucleus. EC decay to the ground state of the 
daughter nucleus is energetically allowed but strongly forbidden by 
weak-interaction selection rules. See text for details. 
}
\label{fig:1}
\end{figure*}

{\textit{Resonant transitions and neutrino energies---}}There are essentially 
two classes of RIOEC transitions: (A) induced EC on nuclei that 
are stable with respect to spontaneous EC due to insufficient energy, with the  
missing energy supplied by the incident $\bar{\nu}_e$; and (B) induced 
EC on nuclei for which spontaneous EC decay is energetically allowed, but the 
transitions to the ground state of the daughter nucleus are strongly 
suppressed by weak-interaction selection rules. Induced transitions to 
excited nuclear states may then still be allowed.  
These two cases are shown schematically in the left and right panels 
of Fig.~\ref{fig:1}, respectively. 

In both cases, the resonance energies are given by 
\be
E_R=-Q_\varepsilon+E_x+E_b\,.
\label{eq:Eres1}
\ee
Here $Q_\varepsilon\equiv M(Z,A) - M(Z-1,A)$ 
\footnote{We use the natural units $\hbar=1=c$ throughout this Letter.}
is the difference of the atomic masses of the initial- and final-state atoms, 
$E_b$ is the excitation energy of the daughter atom (which essentially 
coincides with the binding energy of the missing electron in the daughter 
atom), and $E_x$ is the nuclear excitation energy of the 
daughter nucleus. For atoms that are stable with respect to EC, 
$Q_{\rm EC}\equiv Q_\varepsilon-E_b<0$. 

For the resonance to occur, the resonance condition on the neutrino energy $E$ 
must be satisfied to high accuracy; possible deviations of $E$ from $E_R$ 
are limited by the width of the resonance, which includes 
the natural linewidth of the final state as well as Doppler and 
pressure broadening. 
For the cases of interest to us, the resonance widths are dominated 
by the natural linewidths $\Gamma$ of the 
daughter system and are very small, at the eV 
or sub-eV level. The RIOEC cross section can then be written as 
\cite{Akhmedov:2008jn,Lusignoli:2010eq} 
\be
\sigma=B_0\frac{\Gamma/2\pi}{(E-E_R)^2+\Gamma^2/4}\,.
\label{eq:BW}
\ee
We will concentrate here on allowed $\beta$ transitions, i.e.\ those of 
Fermi (F), Gamow-Teller (GT) and mixed F-GT types; 
the quantity $B_0$ then does not depend on neutrino energy and is given by 
\cite{Akhmedov:2008jn} 
\be
B_0=2\pi G_F^2 |V_{ud}|^2\left\{|M_{\rm F}|^2+g_A^2|M_{\rm GT}|^2\right\}
\left|\Psi_e(R)\right|^2 \kappa\,.
\label{eq:B0}
\ee
Here $G_F$ and $V_{ud}$ are the Fermi constant and the $ud$ element of the 
CKM matrix, respectively, $M_{\rm F}$ and $M_{\rm GT}$ are the Fermi and 
Gamow-Teller nuclear matrix elements, $g_A\simeq 1.276$ is the axial-vector 
coupling constant, $\Psi_e(R)$ is the value of the wave function of the 
captured electron at the nuclear radius, and $\kappa$ accounts for the so-called 
exchange and overlap corrections. The atomic parameters $|\Psi_e(R)|^2$, 
$\kappa$, and $E_b$ also enter the expressions for the rates of spontaneous EC,
and there exist extensive numerical tables of their values   
{\cite{Bambynek:1977zz, EADL_EPICS2025, Schoonjans2011Xraylib, BehrensJanecke1969}}.

The resonance condition for RIOEC may appear to require extreme fine-tuning 
of the antineutrino energy. This is, however, not the case 
when the incoming neutrinos have a continuous energy spectrum extending over 
a range that includes the resonance energy. 
The resonance condition is then always satisfied, but only for 
neutrinos whose energies lie in a narrow interval of width $\sim\Gamma$ around 
$E=E_R$ (see Fig.~\ref{fig:continuous}). While this corresponds 
to a very small fraction of the total antineutrino flux, the cross section 
(\ref{eq:BW}) is proportional to $1/\Gamma$ at the resonance; this resonant 
enhancement compensates for the small fraction of 
detectable neutrinos. As a result, there 
is no strong suppression of the signal, 
but also no particular enhancement of the kind one might expect for 
a resonant process. As demonstrated below, the advantage of RIOEC over 
other $\bar{\nu}_e$ detection methods is not an increased rate, but 
rather that it in principle
allows the detection of neutrinos of unprecedentedly low energies.    
\begin{figure}[h]
\centering
\includegraphics[width=8.2cm,height=4.0cm]{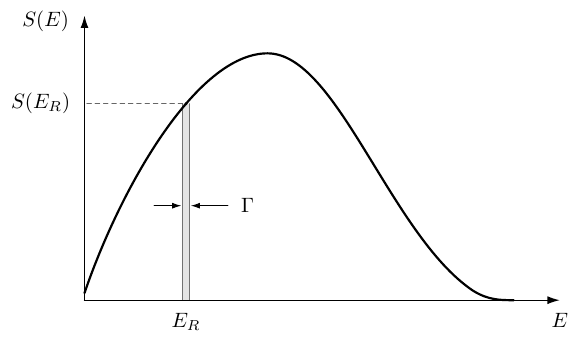}
\caption{
Resonance energy and width in the case of a $\bar{\nu}_e$ source 
with continuous spectrum $S(E)$ (schematic representation).
}
\label{fig:continuous}
\end{figure}

For all neutrino sources of interest except C$\nu$B, the resonance energies 
are very large compared with the resonance widths $\Gamma$; the resonant 
Lorentzian factor multiplying $B_0$ in Eq.~(\ref{eq:BW}) can then be replaced,
to high accuracy, by $\delta(E-E_R)$. This means that the RIOEC cross 
sections are essentially $\Gamma$-independent 
in the limit $\Gamma \ll E_R$. 

Cross sections that are proportional to a $\delta$-function (or $\delta$-like 
functions) are not very informative by themselves; more 
relevant quantities are the effective cross sections 
$\bar{\sigma}$ 
obtained by averaging 
$\sigma(E)$ over the spectrum of the incident antineutrinos. For a normalized 
continuous spectrum $S(E)$%
\footnote{For reactor neutrinos, the spectrum is usually normalized 
to the number of $\bar{\nu}_e$ produced per fission. For other neutrino 
sources, $S(E)$ can be normalized either to the total flux or to unity. 
In the latter case the overall flux enters the expression for the 
event rate as a separate factor. See the End Matter for more details.}, from Eq.~(\ref{eq:BW}) one finds, in the 
limit $\Gamma\ll E_R$,  
\be 
\bar{\sigma}\equiv\int \sigma(E)S(E) dE \simeq B_0\,S(E_R)\,.
\label{eq:sigmaBar}
\ee
This result reveals a crucial property of RIOEC: 

{\textsl{
For allowed $\beta$ transitions ($B_0=\text{const.}$), the effective 
$\bar{\nu}_e$ 
detection cross sections are independent of the neutrino energy and depend 
only on the spectral intensity of the neutrino flux at the resonance, 
$S(E_R)$}.} 

This opens up the possibility of detecting neutrinos of very low energies, 
provided that target nuclides exist that allow transitions with 
sufficiently low resonance energies $E_R$. 

{\textit{Antineutrino detection and  
sources of interest---}}The most widely used 
means of electron antineutrino detection is the inverse $\beta$ decay (IBD) 
reaction on protons, $\bar{\nu}_e+p\to n+e^+$, which has the energy 
threshold $E_{\rm thr}=1.806$ \,MeV. Antineutrinos of lower energies 
can be observed through elastic scattering off electrons, nucleons, or nuclei, 
including coherent elastic neutrino-nucleus scattering (CEvNS).  
These processes have zero physical threshold, but the detection of low-energy 
$\bar{\nu}_e$ is limited by the 
necessity of detecting low-energy recoils. The latter is complicated by 
backgrounds, which rapidly increase with decreasing recoil energy. 
In addition, since 
at energies well below the mass of the target particle the 
scattering cross sections scale as $E^2$, 
the contributions of low-$E$ events are suppressed 
relative to those from higher-energy neutrinos and cannot be reliably  
isolated. 

In contrast, RIOEC is highly selective and measures the $\bar{\nu}_e$ flux 
at essentially discrete values of $\bar{\nu}_e$ energy. Low-energy 
backgrounds can be strongly suppressed through the delayed coincidence 
measurements, as we discuss below.

There are several continuous-spectrum 
sources of electron antineutrinos with important
parts of the spectrum lying at very low energies.

(1) Nuclear reactors. Reactor \(\bar{\nu}_e\) have so far been studied predominantly through IBD on free protons. Their energy spectrum, however, extends well below the threshold of this reaction and is predicted to peak in the \(200\text{--}300\,\mathrm{keV}\) region 
\cite{Kopeikin:2012zz,Perisse:2023efm,Onillon:2021cevns}, see Fig.~\ref{fig:reactor1}. 
\begin{figure*}[t]
\centering
\includegraphics[width=17.8cm]{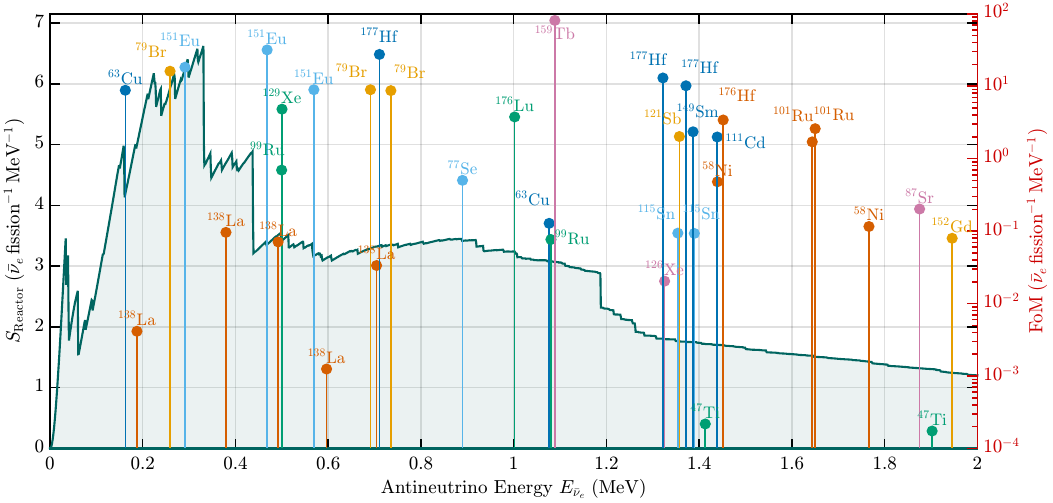}
\caption{
Reactor neutrino spectrum below 2~MeV (as calculated in 
\cite{Perisse:2023efm,Onillon:2021cevns}), with resonance energies for a number of RIOEC 
candidate nuclides superimposed. The right-hand logarithmic scale shows the figure 
of merit  (see text for details). }
\label{fig:reactor1}
\end{figure*}

This low-energy part of the spectrum, which contains nearly 75\% of the integrated antineutrino flux, remains virtually unexplored. Notably, the spectral intensity of the flux at its maximum is predicted to be approximately a factor of 25 larger than in the region $E\sim 4$\,MeV  that gives the dominant contribution to the IBD event rate. RIOEC will allow the low-energy part of the reactor neutrino spectrum to be studied at a number of discrete points. 
Due to their large, controlled  flux, reactor neutrinos are the most promising candidate for the first observation of RIOEC. 

(2) Geoneutrinos. Decay of $\beta$-radioactive elements inside the Earth 
mostly produces electron antineutrinos 
\cite{Smirnov:2019pnj,Ludhova:2013hna,Vitagliano:2019yzm}.
Geoneutrinos have so far been detected only through IBD on
protons, leaving the flux below its 1.8\,MeV threshold unexplored. This
includes an important contribution expected from ${\rm ^{40}K}$ decay 
($E_{\rm max}=1.311$\,MeV), whose measurement would constrain the
potassium content of the Earth's crust and mantle, its heat production by
radioactive decay, and thus models of the Earth's
composition~\cite{LiquidOConsortium:2023bqe}. RIOEC could probe potassium geoneutrinos 
using a low-resonance-energy target. 

(3) Thermal solar neutrinos. Thermal processes in the Sun are expected to 
produce $\nu\bar{\nu}$ pairs of all flavors with keV and sub-keV 
energies, with the spectral maximum in the $\sim$10\,eV\,--\,1\,keV 
range \cite{Haxton:2000xb,Vitagliano:2017odj,Vitagliano:2019yzm}. 
Detecting these low-energy neutrinos through the usual scattering 
processes appears hopeless; however, RIOEC will allow  
detecting 
their $\bar{\nu}_e$ fraction if target 
nuclides with sufficiently low $E_R$ exist.  

(4) C$\nu$B. 
Because of the extremely small energies of relic neutrinos  
(sub-meV range), 
the RIOEC resonance condition would 
require 
the energy of the daughter atom (including possible atomic and 
nuclear excitations) to exceed the 
parent atom's energy by only a tiny amount 
\cite{Cocco:2009rh,Lusignoli:2010eq,Vergados:2013qpa,Lee:2018boo}. 
No such pairs of atoms have been found to date, and it is unclear how 
such minuscule energy differences could be measured experimentally
with sufficient accuracy. 

{\textit{Selection of target candidates. Expected event rates---}}We have 
performed a systematic search for suitable RIOEC target nuclide 
candidates by scanning the nuclear and atomic databases: ENSDF~\cite{ENSDF2026}, 
AME2020~\cite{AME2020II}, NIST~\cite{NIST_IsotopicCompositions}, and EADL~\cite{EADL_EPICS2025}. 
The candidates were required to satisfy the following conditions: 
(i) be stable or very long lived (half-life $T_{1/2}> 1$\,Myr); 
(ii) be naturally abundant; 
(iii) allow resonance energies $E_R<2$\,MeV; 
(iv) have well-established spin/parity assignments and admit allowed
$\beta$ transitions to excited state(s) of the daughter nucleus;
(v) allow RIOEC with electron capture from the $K$-shell;
(vi) allow transitions to excited final states that permit a delayed 
coincidence signature assuming a detector time resolution of $\Delta t =10$~ns. 

We have identified 21 candidate nuclides that allow
37 class (A) RIOEC transitions (several nuclides permit transitions to more than one excited state of the daughter nucleus). 
For class (B) transitions we have found only four potential candidates, three of which, however, are not robust because 
their spin/parity assignments are not reliably established. It is expected that more class (B) candidates will be 
found if some of our search criteria are relaxed. 

To implement the delayed coincidence approach to RIOEC detection, 
we introduce the coincidence probability $P_{\rm coinc}(\Delta t)$ defined as the probability that 
the nuclear de-excitation cascade is still ongoing at the time $\Delta t$ (see the End Matter for details). 
Our delayed coincidence approach allows 
strong suppression of accidental backgrounds; quantitative statements cannot, however, be 
made without considering the details of concrete experiments, which is beyond the scope of this Letter. 
   
For our preliminary selection of suitable targets and estimates
of the expected event rates, we introduced the following approximations
in calculating the RIOEC cross sections.
First, we assumed the squared nuclear matrix elements of the
process to be equal to 0.1 for all the candidate nuclei and all transitions.
This corresponds to $\log(ft)\simeq 4.8$, which is consistent with the typical
range $\log(ft)\simeq  4-6$
for allowed $\beta$ transitions;
the chosen value is about a factor of 60 smaller than the squared matrix
elements of neutron and tritium $\beta$ decays.

Next, we set the factor $\kappa$ that accounts for the exchange and overlap
corrections to unity. This is a very good approximation for $K$-shell
capture and $Z>20$, where the deviation of $\kappa$ from 1 does not exceed
3\% \cite{Bambynek:1977zz}.

To rank the target candidates, we introduce the following figure of merit (FoM): 
\be
{\rm FoM}=(Z/20)^3 S(E_R)f_{\rm nat}P_{\rm coinc}\,.
\label{eq:FoM}
\ee
Here $Z$ is the charge of the parent nucleus, $f_{\rm nat}$ is the natural abundance of the 
candidate isotope, and we have taken into account that to a good approximation 
the electron density $|\Psi_e(R)|^2$ scales with $Z$ as $Z^3$.
Fig.~\ref{fig:reactor1} displays the resonance energies and FoM values for 
all selected class (A) RIOEC candidates 
 and a single robust class (B) candidate ${\rm ^{176}Lu}$, superimposed on the reactor $\bar{\nu}_e$ spectrum 
 for $E<2$\,MeV. The properties of several representative candidates  
are summarized in Table~\ref{tab:candidates}. Of special interest is ${\rm ^{129}Xe}$, because of the 
significant experience with liquid xenon detectors accumulated  in the experimental community. 
The de-excitation signature appears as a 33.2~keV atomic relaxation and a 250~keV dominant  
prompt transition to the metastable level at 28~keV. The half-life of the latter (16.8\,ns) 
yields a reasonable probability to observe a delayed signal, mainly consisting of IC electrons. 
Another particularly attractive RIOEC target is
${\rm ^{63}Cu}$, which combines a favorable FoM, a distinctive
de-excitation signature of ${\rm ^{63}Ni}^{*}$, and a natural isotopic
abundance of 69\%. 
In ${\rm ^{63}Cu}$, the dominant lower-energy resonance drives the emission of 8.3~keV in the 
form of $X$-ray/Auger radiation and a delayed 87.2~keV transition to the ground state  
from a metastable level with a 1.69~$\mu$s half-life. The ready availability of copper could facilitate
the deployment of kiloton-scale targets potentially needed for 
low-$E$ geoneutrino detection. 

Table~\ref{tab:candidates} also presents the expected event 
rates for reactor neutrino experiments. 
The complete list of selected RIOEC candidates  
and a dedicated study of the potential of RIOEC for detecting low-energy 
reactor neutrinos  
will be published elsewhere \cite{ALM2}. 

\begin{table*}[t]
\centering
\begin{minipage}{0.98\textwidth}
\centering
\renewcommand{\arraystretch}{1.25}
\setlength{\tabcolsep}{3pt}
\scriptsize
\resizebox{\linewidth}{!}{
\begin{tabular}{@{}l c c | l c c c c | c  c  c@{}}
\toprule
\shortstack{\textbf{Target}\\{}} & 
\shortstack{\textbf{Abund.}\\{}(\%)} & 
\shortstack{$\boldsymbol{T_{1/2}}$\\{}} & 
\shortstack{\textbf{Daughter}\\{}} & 
\shortstack{$\boldsymbol{E_b}$\\{}(keV)} & 
\shortstack{$\boldsymbol{E_x}$\\{}(keV)} & 
\shortstack{$\boldsymbol{E_R}$\\{}(keV)} & 
\shortstack{$\boldsymbol{P_{\mathrm{coinc}}}$\\{}(\%)} & 
\shortstack{$\boldsymbol{\overline\sigma}$\\{}($\mathrm{cm}^2\mathrm{fiss.}^{-1}$)} & 
\shortstack{$\boldsymbol{N^{\mathrm{nat}}}$\\{}($\mathrm{yr}^{-1}$)} & 
\shortstack{$\boldsymbol{N^{100\%}}$\\{}($\mathrm{yr}^{-1}$)} \\
\midrule
\shortstack{$\nuclide{159}{Tb}(3/2^{+})$} & $100$ & $\mathrm{stable}$ & \shortstack{$\nuclide{159}{Gd}^{*}(5/2^{+})$} & $50.2$ & $67.829(24)$ & $1088.8(7)$ & $76.8$ & $1.3\times 10^{-45}$ & $52.5$ & $52.5$ \\
\midrule
\shortstack{$\nuclide{151}{Eu}(5/2^{+})$} & $47.8$ & $\geq1.7~\mathrm{Eyr}$ & \shortstack{$\nuclide{151}{Sm}^{*}(5/2^{+})$} & $46.8$ & $167.750(7)$ & $291.2(5)$ & $19.6$ & $2.3\times 10^{-45}$ & $11.7$ & $24.6$ \\
\shortstack{$\nuclide{151}{Eu}(5/2^{+})$} & $47.8$ & $\geq1.7~\mathrm{Eyr}$ & \shortstack{$\nuclide{151}{Sm}^{*}(3/2^{+})$} & $46.8$ & $344.909(6)$ & $468.4(5)$ & $63.2$ & $1.2\times 10^{-45}$ & $20.3$ & $42.7$ \\
\shortstack{$\nuclide{151}{Eu}(5/2^{+})$} & $47.8$ & $\geq1.7~\mathrm{Eyr}$ & \shortstack{$\nuclide{151}{Sm}^{*}(5/2^{+})$} & $46.8$ & $445.680(10)$ & $569.1(5)$ & $18.9$ & $1.2\times 10^{-45}$ & $5.7$ & $12$ \\
\midrule
\shortstack{$\nuclide{177}{Hf}(7/2^{-})$} & $18.6$ & $\mathrm{stable}$ & \shortstack{$\nuclide{177}{Lu}^{*}(9/2^{-})$} & $63.3$ & $150.3984(10)$ & $710.6(8)$ & $94.9$ & $2.5\times 10^{-45}$ & $20$ & $108.6$ \\
\shortstack{$\nuclide{177}{Hf}(7/2^{-})$} & $18.6$ & $\mathrm{stable}$ & \shortstack{$\nuclide{177}{Lu}^{*}(5/2^{-})$} & $63.3$ & $761.7063(14)$ & $1321.9(8)$ & $83.1$ & $1.3\times 10^{-45}$ & $9.5$ & $51.4$ \\
\shortstack{$\nuclide{177}{Hf}(7/2^{-})$} & $18.6$ & $\mathrm{stable}$ & \shortstack{$\nuclide{177}{Lu}^{*}(9/2^{-})$} & $63.3$ & $811.4396(22)$ & $1371.6(8)$ & $66.6$ & $1.3\times 10^{-45}$ & $7.4$ & $40.1$ \\
\midrule
\shortstack{$\nuclide{63}{Cu}(3/2^{-})$} & $69.2$ & $\mathrm{stable}$ & \shortstack{$\nuclide{63}{Ni}^{*}(5/2^{-})$} & $8.3$ & $87.220(28)$ & $162.53(3)$ & $99.6$ & $6.7\times 10^{-47}$ & $5.9$ & $8.6$ \\
\midrule
\shortstack{$\nuclide{129}{Xe}(1/2^{+})$} & $26.4$ & $\mathrm{stable}$ & \shortstack{$\nuclide{129}{I}^{*}(3/2^{+})$} & $33.2$ & $278.381(23)$ & $500.4(32)$ & $27.1$ & $6.2\times 10^{-46}$ & $2.7$ & $10.6$ \\
\midrule
\shortstack{$\nuclide{176}{Lu}(7^{-})$} & $2.6$ & $37.6(7)~\mathrm{Gyr}$ & \shortstack{$\nuclide{176}{Yb}^{*}(8^{-})$} & $61.3$ & $1049.8(6)$ & $1002.1(14)$ & $100$ & $2.2\times 10^{-45}$ & $2.7$ & $103.3$ \\
\midrule
\shortstack{$\nuclide{149}{Sm}(7/2^{-})$} & $13.8$ & $\mathrm{stable}$ & \shortstack{$\nuclide{149}{Pm}^{*}(7/2^{-})$} & $45.2$ & $270.170(5)$ & $1386.9(19)$ & $32.6$ & $6.0\times 10^{-46}$ & $1.5$ & $10.8$ \\
\bottomrule 
\end{tabular}
}
\end{minipage}
\caption{Some RIOEC candidate target nuclei. First
and fourth columns show parent nucleus (ground state) and excited
state of daughter nucleus, including spin/parity assignments. Columns 2, 3
and 5--8 present natural abundance of target isotope, 
its half-life, binding energy $E_b$ of missing electron, excitation energy $E_x$ of daughter nucleus, resonance neutrino energy $E_R$, 
and delayed coincidence probability $P_{\rm coinc}$. 
Uncertainty of $E_R$ is typically dominated by that of $Q_\epsilon$ (not shown in the table). 
Last three columns report effective cross section per fission for reactor neutrinos and  
expected numbers of events per year per 100~kg of target in a 
reactor experiment at 15\,m from a 4\,GW$_{\rm th}$ reactor, for natural target
abundance (column 10) and assuming 100\% enrichment (column 11). Possible neutrino oscillations effects are not included 
in expected event numbers.
}
\label{tab:candidates}
\end{table*}

As can be seen from Table~\ref{tab:candidates}, the detection of low-energy reactor neutrinos
via RIOEC can be achieved with detectors with a few hundred kg to a few tons   
of target material. 
Taking into account the spectral intensities and fluxes of 
sub-IBD geoneutrinos \cite{Smirnov:2019pnj,Ludhova:2013hna,Vitagliano:2019yzm}
and of sub-keV thermal solar neutrinos \cite{Haxton:2000xb,
Vitagliano:2017odj,Vitagliano:2019yzm}, one concludes that their detection 
via RIOEC may require detectors with a kiloton to a few tens of kilotons  
of target material. 

Several points warrant discussion. In this Letter, we have confined our 
analysis to order-of-magnitude estimates of the RIOEC cross sections; 
subsequent work must refine these calculations. In particular, the 
relevant nuclear matrix elements must be explicitly evaluated rather 
than uniformly set to a fixed value. 
One needs the matrix elements of transitions between the ground states 
of parent nuclei and the excited states of the daughter ones, which are not 
directly available from experiment, 
unlike in the case of ground-state to 
ground-state transitions. 
One possibility is to extract the relevant matrix 
elements from existing data on nuclear charge exchange reactions; for 
transitions for which such data are not available, one would need to resort 
to theoretical calculations of the nuclear matrix elements. 

All the low-$E_R$ RIOEC candidates we have identified are also candidates for 
the IBD reaction  
$\bar{\nu}_e+(Z,A)\to (Z-1,A)+e^+$ 
with low threshold energies 
$E_{\rm thr}=E_R-E_b+2m_e$. They are sensitive, however, to 
$\bar{\nu}_e$ energies that are larger than those probed by RIOEC by about 
1\,MeV and do not have energy selectivity. The IBD transitions result in 
the same excited states of the daughter nuclei as RIOEC, but they can be 
easily distinguished from the latter because they do not lead to  the 
characteristic atomic excitations and instead produce positrons, whose annihilation 
can be readily detected. Since the IBD reactions depend on the same  nuclear matrix 
elements as the corresponding RIOEC transitions, they can provide direct  
experimental information on these matrix elements. 

We have not discussed RIOEC for $\bar{\nu}_e$ sources with spectral maxima  
well above 2\,MeV (accelerator, atmospheric, and supernova neutrinos), since 
in these cases the usual IBD on protons is very efficient. However, even for 
such sources RIOEC might be useful for studying the parts of the spectra  
that are below 1.8\,MeV.

{\textit{Neutrino oscillations---}}RIOEC may have important applications to neutrino oscillation studies. 
Because the oscillation length is proportional to the neutrino energy, the ability to detect  
lower-energy neutrinos allows 
oscillation experiments to be carried out at shorter distances from the 
neutrino source, 
implying larger 
fluxes at the detector site. As an example, reactor experiments 
aiming to probe the $\Delta m_{31}^2$--$\theta_{13}$ sector 
of the oscillation parameters usually require baselines $L\sim 1$\,km. 
Going from the typical $\bar{\nu}_e$ energies of $\sim 4$\,MeV 
probed by IBD on protons to $\sim 160$\,keV, which is close to the peak 
of the reactor spectrum and can be probed via RIOEC with a 
${\rm ^{63}Cu}$ target, would allow reducing the baseline from 1\,km  to about 40\,m,  
thereby increasing the flux at the detector by more than a factor of 600. 

For a fixed experimental baseline, detecting lower-energy neutrinos 
allows smaller values of neutrino mass-squared differences to be probed.  
In particular, in short-baseline reactor neutrino experiments searching 
for eV-scale sterile neutrinos, the distances to the reactor core are 
between a few meters and a few tens of meters and cannot be reduced much 
further. RIOEC-based oscillation experiments at the same baseline would allow 
probing mass-squared differences as small as $\Delta m_{41}^2\sim 4\times 
10^{-2}$\,eV$^2$ instead of $\sim 1$\,eV$^2$.  

The fact that only neutrinos within very narrow energy intervals can be 
detected via RIOEC means that, for the purposes of neutrino detection, the 
continuous-spectrum sources can be regarded as sources of practically 
monoenergetic neutrinos. This has important implications for neutrino oscillation experiments. 
A spread in neutrino energies leads to the corresponding spread in the 
oscillation lengths, which complicates the analysis and interpretation of 
the data. Monoenergetic neutrinos are therefore  
preferable for oscillation experiments.  
RIOEC thus combines the advantages of continuous spectra of 
the sources (requiring no fine-tuning of neutrino energies to satisfy 
the resonance conditions) with the monoenergetic nature of the detected 
neutrinos. 

For target nuclei for which RIOEC can  
populate several excited states of the daughter nucleus, a single experiment 
will detect neutrinos of several discrete energies, 
thereby simultaneously probing several oscillation lengths. 
In the absence of matter effects (which are negligible in the 
cases under study), the oscillation probabilities 
depend on the neutrino energy $E$ and the experimental baseline $L$ 
through the ratio $L/E$;   
thus, detecting neutrinos of multiple energies amounts to 
probing the oscillations at several baselines.   
Achieving this with conventional neutrino detection methods would require 
multiple detectors or a single movable detector. 
Probing several oscillation lengths in a RIOEC experiment with a single 
fixed detector, however, would require accurate  
knowledge of the neutrino source spectrum and the nuclear matrix 
elements of the corresponding $\beta$-transitions (or their ratios).  
Whether these conditions can be met in practice 
with sufficient accuracy requires further investigation. 
 
{\textit{Applications to reactor monitoring and safeguards---}} 
The capability to detect very low-energy $\bar{\nu}_e$  
implies that RIOEC may have important applications in reactor monitoring. 
In particular, plutonium-producing breeder reactors produce 
${\rm ^{239}U}$ and ${\rm ^{239}Np}$,  which undergo $\beta$ decay  and 
emit $\bar{\nu}_e$ with energies below 1.26\,MeV. Such 
antineutrinos are not accessible to the usual IBD, but can be detected 
via RIOEC; a more detailed discussion 
is given in \cite{ALM2}. 

To summarize, we have proposed a novel approach to the detection of electron 
antineutrinos based on a resonant capture mechanism and a delayed coincidence 
signature. The underlying process, RIOEC, enables the detection of 
$\bar{\nu}_e$ of previously unprobed low energies. We have identified a number 
of candidate nuclides that allow resonant capture of $\bar{\nu}_e$ with 
energies below 2\,MeV and discussed the potential implications 
for studies of reactor neutrinos, 
geoneutrinos, and sub-keV thermal solar neutrinos. 
We also outlined the applications of RIOEC to neutrino oscillation 
experiments and to reactor monitoring and safeguards.  

{\it Acknowledgements.} 
The authors are grateful to Alexei Smirnov and Anthony Onillon for useful
discussions and to Yuri Efremenko for helpful comments on the first version of the manuscript.
L.M. acknowledges support by the International Max Planck Research School for Precision Tests of 
Fundamental Symmetries. 

\bibliography{rioecbib.bib}
\bibliographystyle{apsrev4-1}

\newpage
\maketitle
\onecolumngrid
\clearpage
\begin{center}
\textbf{\large
End Matter}
\vspace{0.1in}
\end{center}
\setcounter{equation}{0}
\setcounter{figure}{0}
\setcounter{section}{0}
\setcounter{table}{0}
\makeatletter
\renewcommand{\theequation}{A\arabic{equation}}
\renewcommand{\thefigure}{A\arabic{figure}}
\renewcommand{\thetable}{S\arabic{table}}
\twocolumngrid
\renewcommand{\thetable}{A\arabic{table}}

We present here the details of our calculations of the effective cross section and event rates as well as our normalization conventions. 
The effective cross section $\bar{\sigma}$ of RIOEC 
is given by $\bar{\sigma}=B_0 S(E_R)$ [Eq.~(\ref{eq:sigmaBar})], with $B_0$ defined in Eq.~(\ref{eq:B0}).
It can be conveniently rewritten as 
\be
\begin{split}
B_0={}&3.16\times 10^{-43}\,{\rm cm^2\,MeV}\,
\frac{|\Psi_e(R)|^2}{{\rm MeV}^3}
\\ &\times\{|M_{\rm F}|^2+g_A^2|M_{\rm GT}|^2\}\kappa\,.
\end{split}
\label{eq:B02}
\ee
In our estimates 
we set the squared nuclear matrix
element of the process  $\{|M_{\rm F}|^2+g_A^2|M_{\rm GT}|^2\}$ equal to 0.1 for all candidate nuclides. 
As discussed in the main text, we set the parameter $\kappa$ that accounts for the atomic exchange and overlap corrections to unity. 
The electron densities at the nuclear radius, $|\Psi_e(R)|^2$, are taken from Ref.~\cite{BehrensJanecke1969}, which tabulates numerical 
results from Dirac--Hartree--Fock and Dirac--Thomas--Fermi calculations with finite nuclear size taken into account, 
assuming a uniformly charged sphere for nuclear charge distribution. 

Since the proposed RIOEC detection approach requires delayed coincidence detection, the expected event rate per target atom is given by
\be
\dot{N}_{\rm ev}=\Phi(\bar{\nu}_e,L)\bar{\sigma}P_{\rm coinc}(\Delta t) P_{ee}(E_R,L)\,.
\label{eq:eventRate1}
\ee
Here $\Phi(\bar{\nu}_e,L)$ is the antineutrino flux at the detector at a distance $L$ from the neutrino source and
$P_{\rm coinc}(\Delta t)$ is the delayed coincidence probability, which will be defined below. The factor 
$P_{ee}(E_R,L)$ accounts for possible neutrino oscillation effects and is the survival probability of $\bar{\nu}_e$ 
of energy $E_R$ after propagation over a distance $L$. 

We model the detector timing response by a minimum resolvable separation $\Delta t$ between prompt and delayed signals. Since the 
atomic de-excitation is very fast (the lifetimes $\tau_{\rm at}$ are typically of the order
of tens of attoseconds), it can be taken to set the starting time for the nuclear de-excitation cascade. 
A RIOEC delayed-coincidence signal then  consists of a prompt signal from atomic relaxation (possibly accompanied by 
unresolved nuclear de-excitations from states with lifetimes $\tau\ll \Delta t$), 
followed by at least one nuclear de-excitation occurring with a delay greater than $\Delta t$. 
$P_{\rm coinc}(\Delta t)$ is thus the probability that the de-excitation cascade is still ongoing at time $\Delta t$, 
so that at least one excited nuclear state decays at a time $t>\Delta t$. It gives the timing acceptance for a minimum delay 
$\Delta t$, assuming unit detection efficiency and no upper coincidence-time cut. 

Let $P_i(t)$ be the probability that the nucleus is in level $i$ at time $t$, given that level $i_0$ is populated at $t=0$. 
The probability that the cascade is still ongoing at time $\Delta t$ is then the sum of the excited-state populations: 
\be
P_{\rm coinc}(\Delta t)=\sum_{i=1}^{i_0}P_i(\Delta t)\,.
\label{eq:Pcoinc1}
\ee
Here the excited states are numbered in order of increasing energy and $i_0$ is the excited nuclear state initially populated
by RIOEC. The label $i=0$ is reserved for the ground state. 

The population probabilities $P_i(t)$ satisfy the rate equations 
\be
\dot{P}_i(t)=\sum_{k=i+1}^{i_0}\lambda_k b_{ki} P_k(t) - \lambda_i P_i(t) 
\label{eq:evol1}
\ee 
with the initial conditions $P_{i_0}(0)=1$ and $P_i(0)=0$ for $0\le i\le i_0-1$. 
Here $\lambda_k=1/\tau_k$ is the total decay rate of level $k$, with mean lifetime $\tau_k$, and $b_{ki}$ is the branching ratio for the transition to level $i$. The branching ratios account for all nuclear de-excitation channels, including IC, and satisfy $\sum_{j=0}^{i-1}b_{ij}=1$. 
Equation~(\ref{eq:evol1}) relates the population probabilities $P_i(t)$ ($i=1,\dots,i_0-1)$ to the populations of the higher-lying states, with $P_{i_0}(t)=e^{-\lambda_{i_0}t}$. 
Reaching the ground state terminates the nuclear de-excitation cascade; subsequent radioactive decay of the daughter ground state, if any, is not included.

Solving the rate equations (\ref{eq:evol1}) for the population probabilities $P_i(t)$  and substituting the result into Eq.~(\ref{eq:Pcoinc1}) at $t=\Delta t$ yields the delayed coincidence probability $P_{\rm coinc}(\Delta t)$.

The normalization convention for the $\bar{\nu}_e$ spectrum $S(E)$ (and, correspondingly, for the RIOEC effective cross section $\bar{\sigma}$) depends 
on the neutrino source under study. For geoneutrinos, calculations usually provide the differential luminosity in units of ${\rm s^{-1}\,MeV^{-1}}$. This 
can be regarded as the total $\bar{\nu}_e$ production rate $\Phi_0$ (in ${\rm s^{-1}}$) multiplied by a spectrum $S(E)$ (in ${\rm MeV^{-1}}$) normalized to unity, or equivalently 
as $S(E)$ normalized to the total number of $\bar{\nu}_e$ produced per second. The flux at the detector depends on the detector location and on the distribution of 
$\beta$-radioactive isotopes in the Earth. For thermal solar neutrinos, calculations produce the expected differential neutrino flux 
at Earth in units of ${\rm cm^{-2}\,s^{-1}\,MeV^{-1}}$ (or ${\rm cm^{-2}\,s^{-1}\,keV^{-1}}$). In our notation this corresponds to the product of the total 
flux $\Phi(\bar{\nu}_e,L)$ and the spectrum $S(E)$ normalized to unity. 

For reactor neutrinos it is customary to normalize the spectrum $S(E)$ (and the detection cross section) to the number of $\bar{\nu}_e$ produced per fission. 
Below the IBD threshold, $\beta$ decays of neutron-activation products can contribute significantly in addition to those of fission products.
The detailed spectrum depends on the fuel composition as well as on the structural materials
and their irradiation history. 
With the per-fission normalization of the spectrum and cross section, the flux $\Phi(\bar{\nu}_e,L)$ in Eq.~(\ref{eq:eventRate1}) should be 
replaced, in the isotropic point-source approximation, by 
$\frac{\dot{N}_{\rm fiss}}{4\pi L^2}$, 
where $\dot{N}_{\rm fiss}$ is the reactor fission rate and $L$ is the
distance from the core center to the detector. 

For our calculations of the cross sections and expected event rates quoted in Table~\ref{tab:candidates}, we evaluated 
$S(E_R)$ using an illustrative spectrum for a commercial pressurized-water reactor~\cite{Perisse:2023efm,Onillon:2021cevns}.
A reactor thermal power $P_{\rm th}=4\,{\rm GW}$ and a thermal energy per fission
$\langle E_{\rm fiss}\rangle\simeq205\,{\rm MeV}$ were adopted. 
At $L=15\,{\rm m}$, these values give 
\(
{\dot{N}_{\rm fiss}}/{4\pi L^2}
\simeq4.3\times10^{12}\,{\rm fissions\,cm^{-2}\,s^{-1}}\,.
\)

\end{document}